\documentclass[runningheads]{llncs}
\usepackage[T1]{fontenc}
\usepackage{graphicx}
\usepackage{listings}
\usepackage{xcolor}
\lstdefinestyle{pythonstyle}{
    language=Python,
    basicstyle=\small\ttfamily,
    numbers=left,
    numberstyle=\tiny,
    numbersep=5pt,
    backgroundcolor=\color{white},
    showstringspaces=false,
    commentstyle=\color{green!60!black},
    keywordstyle=\color{blue},
    stringstyle=\color{red},
    breaklines=true,
    breakatwhitespace=true,
    tabsize=4,
    frame=tb,  % This sets borders on left and right only
    framesep=5pt,
    xleftmargin=5pt,
    xrightmargin=5pt
}
\usepackage{tikz}
\usepackage{tikz-dependency}

\newcommand*\circled[1]{\tikz[baseline=(char.base)]{
            \node[shape=circle,draw,inner sep=.6pt] (char) {#1};}}
            
\begin{document}
\title{\texttt{rerankers}: A Lightweight Python Library to Unify Ranking Methods}
%
%\titlerunning{Abbreviated paper title}
% If the paper title is too long for the running head, you can set
% an abbreviated paper title here
%
\author{Benjamin Clavié}
\authorrunning{B. Clavié}
% First names are abbreviated in the running head.
% If there are more than two authors, 'et al.' is used.
%
\institute{Answer.AI\\
\email{bc@answer.ai}\\
}
\maketitle              % typeset the header of the contribution
\begin{abstract}
This paper presents \texttt{rerankers}, a Python library which provides an easy-to-use interface to the most commonly used re-ranking approaches. Re-ranking is an integral component of many retrieval pipelines; however, there exist numerous approaches to it, relying on different implementation methods. \texttt{rerankers} unifies these methods into a single user-friendly interface, allowing practitioners and researchers alike to explore different methods while only changing a single line of Python code. Moreover, \texttt{rerankers} ensures that its implementations are done with the fewest dependencies possible, and re-uses the original implementation whenever possible, guaranteeing that our simplified interface results in no performance degradation compared to more complex ones. The full source code and list of supported models are updated regularly and available at \url{https://github.com/answerdotai/rerankers}.
\end{abstract}
\section{Introduction}

A common approach to information retrieval is the use of a two-stage retrieval pipelines: a small set of candidate documents is first retrieved by a computationally efficient retrieval method, to then be re-ranked by a stronger, generally neural network-based, model~\cite{reranking}. This method is used in order to mitigate latency constraints: while models frequently used for re-ranking are particularly powerful, as they're aware of both documents and queries at encoding time~\cite{monobert}, this efficiency comes at a prohibitive computational cost~\cite{monot5}. As a result, in order to use them on any larger scale document sets, it is necessary to use more lightweight method, such as BM25~\cite{bm25} or DPR~\cite{dpr}, which do not require such expensive calculation, to create a smaller set of document. This smaller set can then be processed in a short amount of time by the re-ranking model.

This approach has been widely adopted in Information Retrieval systems, consistently yielding stronger results than relying on a single retriever without re-ranking~\cite{rocketqav2,sparsedense}. Moreover, such models are very powerful, as their output scores can be used for knowledge distillation~\cite{gravedistill}, where first-stage retrieval models are trained to emulate the scores generated by re-ranking models. Knowledge Distillation from re-ranking model scores has been consistently shown to allow the training of stronger first-stage retrievers~\cite{jacolbertv2.5,spladev3,rocketqav2,colbertv2}.

Over time, there has been an increasing number of neural re-ranking methods. Most early neural re-ranking models were \textit{cross-encoders} ~\cite{monobert}, framing the task as a binary sentence classification task and using a pre-trained encoder model such as BERT~\cite{bert}. However, many new methods have appeared, such as MonoT5~\cite{monot5}, leveraging sequence-to-sequence models such as T5\~cite{t5} rather than BERT-like models, or ColBERT-based methods, which repurpose late-interaction retrieval mdoels such as ColBERTv2~\cite{colbertv2} as re-ranking models. Other work also explores different paradigms, such as \textbf{listwise} re-ranking using sequence-to-sequence models, where the model is not tasked with outputting a score for a \textit{[query, document]} pairs, but rather, to directly output a list of documents ordered by their relevance to the given query~\cite{lit5}. Work such as FlashRank~\cite{flashrank}, has focused on optimised methods to speed up re-ranking when performed on weaker inference hardware, such as CPUs. Finally, some of the best-performing re-ranking methods, such as Cohere-Rerank~\cite{coherererank}, while observed in literature to be competitive with some of the best models~\cite{rankgpt}, are currently only accessible via an online API.

Some recent approaches have also explored the use of Large Language Models (LLMs) as re-rankers. This has been the topic of a lot of attention, with some methods using very powerful LLMs, such as GPT-4~\cite{gpt4}, to perform zero-shot listwise re-ranking~\cite{rankgpt}, while others purposefully fine-tuning smaller models on the output of larger models, another form of knowledge distillation, to create powerful list-wise re-rankers such as RankZephyr~\cite{rankzephyr}. Very recent work in this space has also explored the use of inference-time model compression techniques, in order to use LLM models such as Gemma2~\cite{gemma2} as re-rankers at a fraction of the computational cost~\cite{bgegemma}.

\subsection{Contribution}

These methods all come with different trade-offs, and may be more or less well-suited to different usecases. However, the multiplication of approach makes keeping up with existing research and evaluating new methods is often needlessly difficult. In some cases, adopting a different method might require substantial code changes, due to the wide variety of dependencies and input/output formats across methods.
This issue also makes it more difficult for newer approaches to reach large adoption, as their exploration represents a cost which can be prohibitive.

To alleviate these issues, we introduce \texttt{rerankers}, a simple, light-weight Python library which seeks to unify re-ranking methods. \texttt{rerankers} relies on three core principles: \circled{1} It must provide an easy way to load, use, and swap in\/out various re-ranking methods, \circled{2} It should be as non-intrusive as possible, without requiring the user to greatly modify their environment or codebase to adopt and \circled{3} It must not result in a ranking performance degradation when compared to existing implementations. The library is compatible with all modern Python versions and provides a simple, user-friendly interface to most common re-ranking approaches. In doing so, it also comes with utility to accommodate various use-cases, such as retrieving only the top-k candidates for a given query, or only outputting scores to use for knowledge distillation.

\section{System Overview}

\subsection{The \texttt{Reranker} class}

Everything within the library is organised around a central class named \texttt{Reranker}. It is used as the main interface to load models, no matter the underlying implementation or requirements. Its interface is demonstrated in Listing~\ref{lst:reranker}.

The \texttt{Reranker} class has a single exposed method, \texttt{rank}, which takes in a query and a set of document and returns a \texttt{RankedResults} object, presented in Section~\ref{sec:rankedresults}, containing the re-ranked documents.

\begin{lstlisting}[style=pythonstyle, caption={Examples of how to construct a Ranker and use it for re-ranking.}, label={lst:reranker}]
# Initialising a BERT-like cross-encoder model
ranker = Reranker(MODEL_NAME_OR_PATH, model_type='cross-encoder')

# MonoT5-based models, with a specified dtype
ranker = Reranker(MODEL_NAME_OR_PATH, model_type = "t5", dtype=torch.float32)

# Flashrank models, with a specified device
ranker = Reranker(MODEL_NAME_OR_PATH, model_type='flashrank', device="cpu")
# ... and so on
# Every Reranker then has a single `rank` method, which performs inference.
results = ranker.rank(query="Who wrote Spirited Away?", docs=["Spirited Away [...] is a 2001 Japanese animated fantasy film written and directed by Hayao Miyazaki. ", "Lorem ipsum..."], doc_ids=[0,1])
\end{lstlisting}

\texttt{rerankers} is fully integrated into the huggingface transformers ecosystem~\cite{transformers}, and can load any model compatible with it directly from the HuggingFace hub, as well as models stored locally.

\subsection{Handling \texttt{RankedResults}}
\label{sec:rankedresults}

Similarly to how \texttt{Reranker} serves as a single interface to various models, \texttt{RankedResults} objects are a centralised way to represent the outputs of various models, themselves containing \texttt{Result} objects.

This class comes with various helper methods, presented in Listing~\ref{lst:results}, which makes result easy to retrieve for common use cases. Both \texttt{RankedResults} and \texttt{Result} are fully transparent, allowing you to iterate through RankedResults and retrieve any of their attributes.

\begin{lstlisting}[style=pythonstyle, caption={A quick overview of RankedResults and its key methods and accessors.}, label={lst:results}]
# Ranking a set of documnets returns a RankedResults object, preserving meta-data and document-ids.
results = ranker.rank(query="I love you", docs=["I hate you", "I really like you"], doc_ids=[0,1], metadata=[{'source': 'twitter'}, {'source': 'reddit'}])
results
> RankedResults(results=[Result(document=Document(text='I really like you', doc_id=1, metadata={'source': 'twitter'}), score=-2.453125, rank=1), Result(document=Document(text='I hate you', doc_id=0, metadata={'source': 'reddit'}), score=-4.14453125, rank=2)], query='I love you', has_scores=True)

# RankedResults comes with various built-in functions for common uses, such as .top_k(), and all attributes are accessible:
results.top_k(1).text
> 'I really like you'

# It's also possible to directly fetch the score given to a single document
results.get_score_by_docid(1)
> -4.14453125
\end{lstlisting}

\texttt{rerankers} allows users to specify document ids and document meta-data, which are preserved post-scoring in the \texttt{RankedResults} object. This object also contains the original text of each document, as well as global attribute, \texttt{has\_scores}, indicating whether or not the Results are simply ordered (by a list-wise re-ranker such as RankGPT~\cite{rankgpt}), or if each document has an individual relevance score.

\subsection{Extensibility and Modularity}

\textbf{Modularity} \texttt{rerankers} is designed specifically with ease of extensibility in mind. All approaches are independently-implemented and have individually-defined sets of dependencies, which users are free to install or not based on their needs. Informative error messages are shown when a user attempts to load a model type that is not supported by their currently installed dependencies.

\textbf{Extensibility} As a result, adding a new method simply requires making its inputs and outputs compatible with the \texttt{rerankers}-defined format, as well as a simple modification of the main \textsc{Reranker} class to specify a default model. This approach to modularity has allowed us to support all the approaches with minimal engineering efforts. We fully encourage researchers to integrate their novel methods into the library and will provide support for those seeking to do so.

\section{Comparison with Existing Tools}

\subsection{Existing Libraries}
As of the making of this work, no library providing a consistent API to access common reranking methods existed, as such, \texttt{rerankers} has no direct equivalent. Extensive information retrieval frameworks such as Terrier~\cite{terrier} and Anserini~\cite{anserini} do integrate re-ranking capabilities, however, these aim to be much more full-featured retrieval platforms, and take a fundamentally different development approach to \texttt{rerankers}' low footprint approach.

For most individual methods integrated within \texttt{rerankers}\footnote{With the exception of ColBERT, as all previous reranking-only public implementations suffered from issues impacting performance}, an existing implementation or way to access the online API already existed. \texttt{rerankers} is not a direct competitor to any of these implementations. In fact, our library often re-uses parts of the original authors' code or acts as a wrapper around their library, if a mature one with moderate dependencies exists~\cite{rankzephyr,rankvicuna,lit5}.

\subsection{Performance}

In order to ensure performance parity with existing implementations, we conduct top-1000 reranking evaluations on three commonly used datasets\footnote{a subset of the MS Marco passage retrieval dataset, as well as Scifact and TREC-Covid, all three being subsets of the BEIR benchmark~\cite{beir}.}.

For most models implemented in the library, over 5 runs, we achieve parity with either the existing implementation code and reported results from the litterature. A notable exception is RankGPT~\cite{rankgpt}, where our results over all runs were noticeably different from the paper's reported results~\footnote{The results we obtained were worse than the official ones in 4 runs, and better in 1.}. We observed similar behaviour when using the author's own code, and believe that this is likely due to the difficulty of reproducing experiments with unreleased, API-only models such as the GPT family~\cite{gptrepro}.

\section{Conclusion}

In this work, we introduced \texttt{rerankers}, a lightweight python library to support the use of various re-ranking methods in various retrieval usecases. \texttt{rerankers} provides a simple, unified interface to using nearly all commonly encountered approaches to re-ranking in a single, lightweight package, without any detriment to performance. We hope that re-rankers will support both practitioners and researchers in the future, by considerably lowering the barrier to entry formerly present for less well-known approaches and providing easy extensibility to implement upcoming re-ranking methods within a familiar interface. 

In the future, we are hoping for the rerankers library to also support fine-tuning, to simplify training and comparing in-domain models within a familiar interface.

Finally, it is worth noting that \texttt{rerankers} has already contributed to scholarly work, by producing the model scores used in distillation~\cite{jacolbertv2.5}, or becoming the authors' way of releasing their new Portuguese re-ranking models~\cite{ptt5v2}.

%
% ---- Bibliography ----
%
% BibTeX users should specify bibliography style 'splncs04'.
% References will then be sorted and formatted in the correct style.
%
% \bibliographystyle{splncs04}
% \bibliography{mybibliography}
%
\bibliographystyle{splncs04}
\bibliography{bib}
\end{document}